# Time Series Analysis of Big Data for Electricity Price and Demand to Find Cyber-Attacks part 2: Decomposition Analysis

1st Mohsen Rakhshandehroo
*Electrical and Computer Engineering*
*Shiraz University*
Shiraz, Iran
mohsenrakhshandehroo@gmail.com

2nd Mohammad Rajabdorri
*Electrical and Computer Engineering*
*Shiraz University*
Shiraz, Iran
Rajabdorri@gmail.com

*Abstract*—In this paper, in following of the first part (which ADF tests using ACI evaluation) has conducted, Time Series (TSs) are analyzed using decomposition analysis. In fact, TSs are composed of four components including trend (long term behaviour or progression of series), cyclic component (non-periodic fluctuation behaviour which are usually long term), seasonal component (periodic fluctuations due to seasonal variations like temperature, weather condition and etc.) and error term. For our case of cyber-attack detection, in this paper, two common ways of TS decomposition are investigated. The first method is additive decomposition and the second is multiplicative method to decompose a TS into its components. After decomposition, the error term is tested using Durbin-Watson and Breusch-Godfrey test to see whether the error follows any predictable pattern, it can be concluded that there is a chance of cyber-attack to the system.

## I. INTRODUCTION

In this paper, to find out that TS errors (or called residual's interchangeably) follows any particular patterns or not and to obtain the residual values of TSs, we conducted two classical methods of TS decomposition and then we analyzed the residual terms of TSs for both decomposition method to find anomaly in residual distributions. To evaluate the TS residual values, Durbin-Watson (DW) and Breusch-Godfrey (BG) test as well as visualization method are applied [1] - [7].

## II. TIME-SERIES DECOMPOSITION

Generally, there are plenty of time-series decomposition methods that have been proposed in the field literature [8] - [16]. Among those methods, there are two classical and well known industrial applicable methods: additive and multiplicative. In following both are described.

### A. Additive Decomposition (AD) of TSs

### A. Additive Decomposition (AD) of TSs

In this model, the TS $y_t$ is considered to be consists of components including trend, seasonal and residual (errors) term. This model can be formulated as follows [17]:

$$y_t = S_t + T_t + R_t \qquad (1)$$

In (1), $S_t$ is the seasonal, $T_t$ is the trend-cycle and $R_t$ is residual components. Before discussing about AD, it must be mentioned that if TS is daily moving average ($m$) is considered to be $m = 7$ and if TS is monthly $m = 12$ and if data is seasonal $m = 4$. Considering aforementioned explanation the AD steps are as follows:

1. If $m$ is an even number, the trend-cycle $\hat{T}$ is $2 \times$ (moving average($m$)) and if $m$ an odd number then $\hat{T} =$ (moving average($m$))
2. De-trend the $T_s$ using $y_t - \hat{T}$
3. To obtain the $\hat{S}_t$ for each season, take the average for the de-trended values for that season. For example, for weekly data, $\hat{S}_t$ for the first week is the average of de-trended values of first weeks in TS.
4. The residual values will be [18]:

$$\hat{R}_t = y_t - \hat{T}_t - \hat{S}_t \qquad (2)$$

### B. Multiplicative Decomposition (MD) of TSs

In this model, TS is considered to be as (3):

$$y_t = S_t \times T_t \times R_t \qquad (3)$$

Like AD method, the corresponding steps for MD are as follows:

1. If $m$ is an even number, the trend-cycle $T$ is $2 \times$ (moving average($m$)) and if $m$ an odd number then $\hat{T} =$ (moving average($m$))
2. De-trend the TS using $\frac{y_t}{\hat{T}_t}$
3. To obtain the $\hat{S}_t$ for each season, take the average for the de-trended values for that season. For example, for weekly data, $\hat{S}_t$ for the first week is the average of de-trended values of first weeks in TS.
4. The residual value is as:

$$\hat{R}_t = \frac{y_t}{\hat{T}_t \hat{S}_t} \qquad (4)$$

## III. SERIAL CORRECTION (AUTO CORRECTION) OF RESIDUAL VALUES

In this section, serial correction (which is also known as error auto-correlation), is analyzed. It must be noted that if the residual values $R_t$ of TS doesn't follow the normal distribution, then there is anomaly in data. To ensure that there is no auto correction in error, there are two well known tests that can be implemented on data [19] - [23]. In following these two are introduced.

### A. Durbin-Watson (DW) Test

One of the most common tests to find out the randomness of error term is DW test. Consider the first order auto-regression model (from the part 1 of this paper), we have $y_t = \alpha_0 + \gamma y_{t-1} + \varepsilon_t$. If the error term is not random, then it is reasonable to assume that there is a correlation between error at the time $t$ and the first lag of errors as follows:

$$\varepsilon_t = \rho \varepsilon_{t-1} + U_t, \quad U_t \sim N(0, \delta^2) :: d \quad (5)$$

Under DW test, the hypothesis is as follows:

$$\begin{array}{l} H_0 : \rho = 0 \\ H_1 : \rho \neq 0 \text{ or } \rho < 0 \end{array} \quad (6)$$

Equation (6) emphasizes that under the Null hypothesize, there is no auto-correlation between errors, however the $H$, shows the auto-correlation. Based on DW test, the DW statistic is defined as follows [24]:

$$DW = \frac{\sum_{t=2}^{n}(\varepsilon_t - \varepsilon_{t-1})^2}{\sum_{t=1}^{n}(\varepsilon_t)^2}$$

$$DW = \quad \sum_{t=1}^{n}(\varepsilon_t)^2 \quad (7)$$

If DW test is less than a critical value, the we reject the $H_0$, which means there is auto-correlation between error term and their first lag and thus, there is possibility of cyber-attacks toward the electricity grid. As an intuition to DW statistic, it can be mentioned that if we have auto-correlation, the next error term would be as similar as the current error term, therefore the $(\varepsilon_t - \varepsilon_{t-1})^2$ would be a small value. In contrast, if the error term is random, $(\varepsilon_t - \varepsilon_{t-1})^2$ in general is a big term. Unlike the common hypothesis tests, there is not a certain critical value for DW statistic [25] - [26]. But instead, there are upper $d_u$ and lower $d_l$ critical values so that:
- DW statistic > $d_u$ → we cannot reject $H_0$
- DW statistic < $d_l$ → we reject $H_0$ and conclude $H_1$
- $d_l < D < d_u$ → we cannot include eighter auto-correlation or randomness.

DW test only checks the first lag auto-correlation and longer lags are not considered. It is proven that DW statistic $\approx 2(1 - \hat{\rho})$ and as a role of thumb it can be concluded:
- for DW statistic greater than 2, there is no auto-correlation (we cannot reject $H_0$
- for DW close to zero, there is perfect auto-correlation in errors.

The auto correlation consequences can be listed as below:
1. Regression coefficients remain unbiased, but are not longer minimum variance estimates.
2. For positive auto correlation MSE estimate of residual variance will underestimate.

No matter which method is used to gain the $\rho$, but DW statistic can be approximated as below (it must be noted that is the most common way):

$$\hat{\rho} = cor(\varepsilon_t, \varepsilon_{t-1})$$

$$DW \subset 2(1 - \hat{\rho}) \quad (8)$$

, there are

Based on (8), and considering $\hat{\rho} = cor(\varepsilon_t, \varepsilon_{t-1})$ three restrict conditions:
1. $\rho = 1 \to DW = 0$ (it means the most positive first order of correlation)
2. $\rho = 0 \to DW = 2$ (It means there is no auto correlation)
3. $\rho = -1 \to DW = 4$ (It means the most first order of auto correlation)

*1) Removing the Auto Correlation from TS:* To remove the auto-correlation of order 1, it needs to transform the TS by the following equation, consider the AR(1) model as:

$$y_t = \beta_0 + \beta_1 x_t + \varepsilon_t \quad (9a)$$

$$\varepsilon_t = \rho \varepsilon_{t-1} + U_t, \quad U_t(0, \delta^2) \quad (9b)$$

Let's define the transformed variables as below ($x_t$ in the TS is $y_{t-1}$):

$$\begin{array}{ll} y_t^j = y_t - \rho y_{t-1} & \beta^j = \beta_0(1-\rho) \\ x_t^j = x_t - \rho x_{t-1} & \beta_1^j = \beta_1 \end{array} \quad (10)$$

be as:
In case of AR(1) model for the TS the transformation would

$$y_t = \beta_0 + \beta_1 y_{t-1} + \varepsilon_t \quad (11)$$

$$y_{t-1} = \beta_0 + \beta_1 y_{t-2} + \varepsilon_{t-1} \quad (12)$$

$$\varepsilon_t = \rho \varepsilon_{t-1} + V_t, V_t \sim (0, \delta^2) \quad (13)$$

Multiplying (12) by $\rho$ and subtracting from (11), we will have:

$$y_t - \rho y_{t-1} = \beta_0(1-\rho) + \beta_1(y_{t-1} - \rho y_{t-2}) + \varepsilon_t - \rho \varepsilon_{t-1} \quad (14)$$

The last towo terms in (14) would be the equal to $U_t$ which follows he normal distribution and we can then run the LS to find the $\beta_0, \hat{\beta_0}, \hat{\beta_1}$ and $U_t$. The flowchart for the DW test and transformed TS is show in Fig. [1]

Considering the DW test there are major drawbacks that can be listed as follows;
1. DW interpretation is hard and complicated
2. $\hat{\rho}$ is inconsistent even for large samples
3. DW only considers the auto correlation of order 1.

Regrading the above-mentioned condition, there is another comprehensive statistical test for auto correlation in the next section.

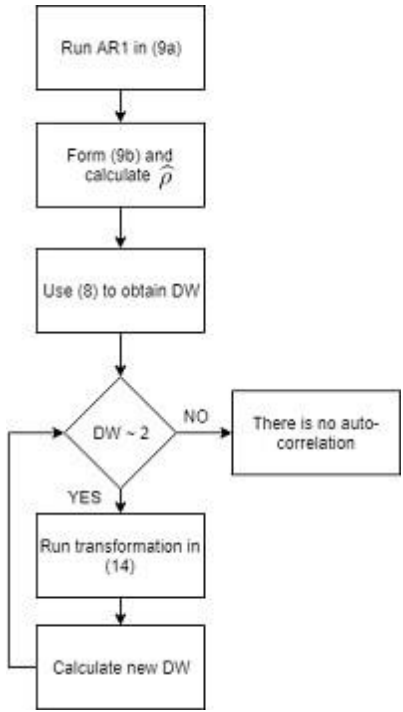

Fig. 1. DW Test and Transformed TS

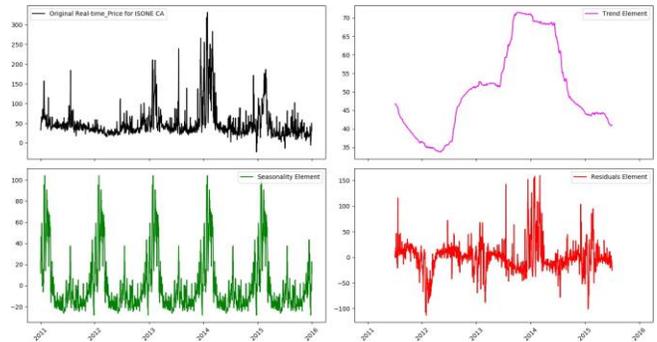

Fig. 2. AD decomposition for the real-time prices for ISONE.

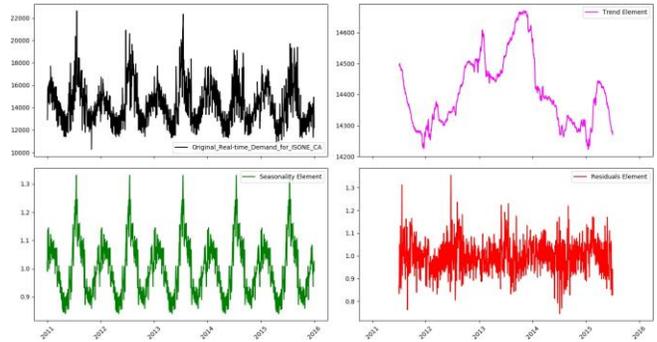

Fig. 3. MD decomposition for the real-time demand for ISONE.

## B. Breusch-Godfrey or (LM) Test

In this model, we use an auxiliary regression equation for modelling the error. To run the LM test, the step by step procedure is as follows:

1) Estimate $y_t = \alpha + \beta x_{t)} + \varepsilon_t$ by ordinary least square method
2) Obtain the residuals $\varepsilon_t$
3) Run an auxiliary regression as follows;

$$\hat{\varepsilon}_t = Y + sx_t + \rho_1 \widehat{\varepsilon_{t-1}} + \rho_2 \widehat{\varepsilon_{t-2}} + \ldots + \rho_k \widehat{\varepsilon_{t-k}} + U_t \quad (15)$$

Since in (15), it is considered that $x_t$ is exogenous regressor, so it doesn't influence on $\rho_1, \ldots, \rho_k$. So the test hypothesis is as follows:

4) Form the test hypothesis as;

$$\begin{array}{l} H_0: \rho_1 = \rho_2 = \ldots = \rho_k = 0 \quad or \quad R^2 = 0 \\ H_1: \rho_i \neq 0 \quad \forall_{i=1,\ldots,k} \quad or \quad R^2 > 0 \end{array} \quad (16)$$

In (16) the Null hypothesis emphasizes that there is no auto-correlation while $H_1$ states that there is some sort of auto-correlation. The Breusch-Godfrey test needs LM statistic which proven to follow $\chi^2_k$ distribution in which $k$ is number of auto-correlation lags that we have considered.

5) Calculate LM statistic as following;

$$LM = nR^2 \sim \chi^2_k, \quad n = T - k \quad (17)$$

T is the total number of samples. If LM statistic is greater than the critical value from $\chi^2_k$ with $k$ degree of freedom, we will reject the $H_0$, so there si auto-correlation of K lag in this case. If LM statistic is less than critical value from $\chi^2_k$ with $k$ degree of freedom, then we cannot reject Null hypothesis and there is no auto-correlation.

As a good rule of thumb choosing the $k$ can be based on seasonality of your data, if the data is seasonal $k = 4$, if daily $k = 7$ and if monthly $k = 12$ are good estimation. However, Akaike information criteria (AIC) can be applied as explained in the first part of this paper.

## IV. RESULTS FOR NEW ENGLAND BIG DATA

In this section, first electricity price and demand for 9 different operational zones of NE are decomposed using both AD and MD methods. Then DW and GB test are applied to these data to find possible auto-correlation as source of cyber-attack to the system.

### A. Decomposition and Auto-Correlation for Zone 1

The AD and MD decomposition for electricity prices and demands are depicted in Figs. [2]-[3]. The test results for DW and GB test are listed in Table. [1].

### B. Decomposition and Auto-Correlation for Zone 2

The AD and MD decomposition for electricity prices and demands are depicted in Figs. [4]-[5]. The test results for DW and GB test are listed in Table. [2].

### C. Decomposition and Auto-Correlation for Zone 3

The AD and MD decomposition for electricity prices and demands are depicted in Figs. [6]-[7]. The test results for DW and GB test are listed in Table. [3].

TABLE I
ISONE CA REAL-TIME PRICE PREDICTION

| RTP Prediction Error | | | | |
|---|---|---|---|---|
| coeff. | std err. | t | [0.025 | 0.975] |
| -0.1009 | 0.023 | -4.331 | -0.147 | -0.055 |
| **F-statistic** | 18.76 | **BIC** | 1.610e+04 | |
| **R-squared** | 0.010 | **Adj. R-squared** | 0.010 | |
| **Log-Likelihood** | -8047.6 | **AIC** | 1.610e+04 | |
| **Prob (Omnibus)** | 0.000 | **Skew** | 0.291 | |
| **Jarque-Bera (JB)** | 9542.915 | **Prob (JB)** | 0.00 | |
| **Prob (F-statistic)** | 1.56e-05 | **Omnibus** | 375.380 | |
| **Kurtosis** | 14.187 | **Cond. No.** | 1.00 | |

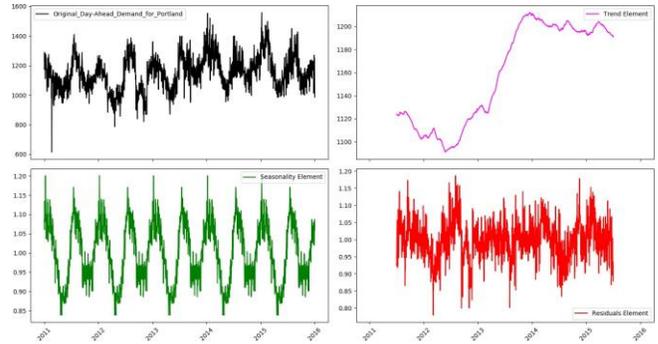

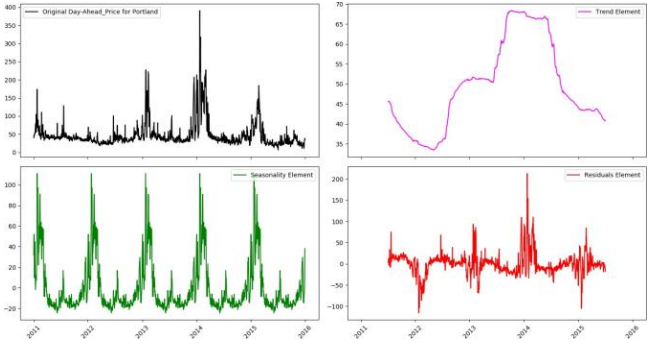

Fig. 4. AD decomposition for the Day-ahead prices for Portland.

Fig. 5. MD decomposition for the Day-ahead prices for Portland.

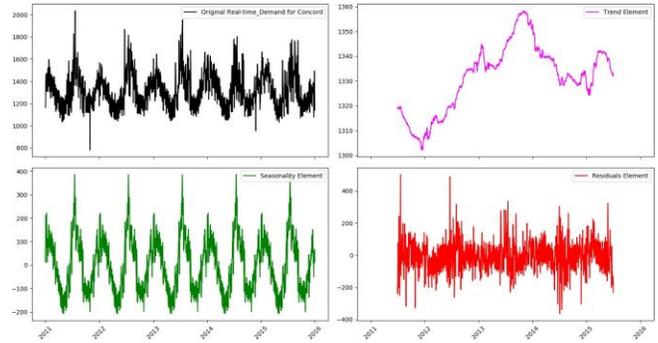

Fig. 6. AD decomposition for the real-time demand for Concord.

### D. Decomposition and Auto-Correlation for Zone 4

The AD and MD decomposition for electricity prices and demands are depicted in Figs. [8]-[9]. The test results for DW and GB test are listed in Table. [4].

### E. Decomposition and Auto-Correlation for Zone 5

The AD and MD decomposition for electricity prices and demands are depicted in Figs. [10]-[11]. The test results for DW and GB test are listed in Table. [5].

### F. Decomposition and Auto-Correlation for Zone 6

The AD and MD decomposition for electricity prices and demands are depicted in Figs. [12]-[13]. The test results for DW and GB test are listed in Table. [6].

### G. Decomposition and Auto-Correlation for Zone 7

The AD and MD decomposition for electricity prices and demands are depicted in Figs. [14]-[15]. The test results for DW and GB test are listed in Table. [7].

TABLE II
PORTLAND DAY-AHEAD PRICE PREDICTION ERROR

| RTP Prediction Error | | | | |
|---|---|---|---|---|
| coeff. | std err. | t | [0.025 | 0.975] |
| 0.1258 | 0.023 | 5.415 | 0.080 | 0.171 |
| **F-statistic** | 29.32 | **BIC** | 1.471e+04 | |
| **R-squared** | 0.016 | **Adj. R-squared** | 0.015 | |
| **Log-Likelihood** | -7352.2 | **AIC** | 1.471e+04 | |
| **Prob (Omnibus)** | 0.000 | **Skew** | -0.012 | |
| **Jarque-Bera (JB)** | 18831.860 | **Prob (JB)** | 0.00 | |
| **Prob (F-statistic)** | 6.93e-08 | **Omnibus** | 415.283 | |
| **Kurtosis** | 18.737 | **Cond. No.** | 1.00 | |

TABLE III
CONCORD RT DEMAND PREDICTION ERROR

| RTP Prediction Error | | | | |
|---|---|---|---|---|
| coeff. | std err. | t | [0.025 | 0.975] |
| 0.1566 | 0.023 | 6.765 | 0.111 | 0.202 |
| **F-statistic** | 45.77 | **BIC** | 2.169e+04 | |
| **R-squared** | 0.024 | **Adj. R-squared** | 0.024 | |
| **Log-Likelihood** | -10843. | **AIC** | 2.169e+04 | |
| **Prob (Omnibus)** | 0.000 | **Skew** | 0.346 | |
| **Jarque-Bera (JB)** | 165.852 | **Prob (JB)** | 9.68e-37 | |
| **Prob (F-statistic)** | 1.79e-11 | **Omnibus** | 87.361 | |
| **Kurtosis** | 4.305 | **Cond. No.** | 1.00 | |

TABLE IV
BURLINGTON RT DEMAND PREDICTION ERROR

| RTP Prediction Error | | | | |
|---|---|---|---|---|
| coeff. | std err. | t | [0.025 | 0.975] |
| 0.1453 | 0.023 | 6.268 | 0.100 | 0.191 |
| **F-statistic** | 39.29 | **BIC** | 2.169e+04 | |
| **R-squared** | 0.021 | **Adj. R-squared** | 0.021 | |
| **Log-Likelihood** | -9123.6 | **AIC** | 1.825e+04 | |
| **Prob (Omnibus)** | 0.000 | **Skew** | 0.315 | |
| **Jarque-Bera (JB)** | 48.753 | **Prob (JB)** | 2.59e-11 | |
| **Prob (F-statistic)** | 4.54e-10 | **Omnibus** | 41.952 | |
| **Kurtosis** | 3.494 | **Cond. No.** | 1.00 | |

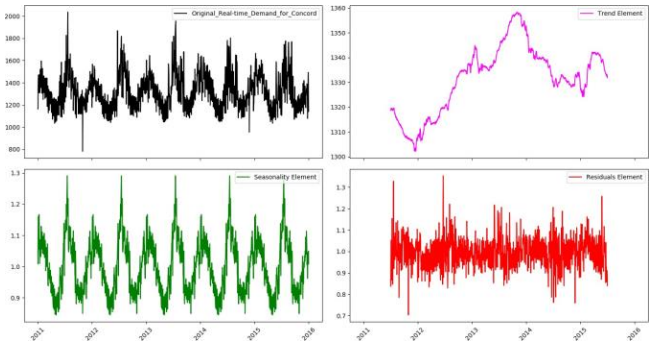

Fig. 7. MD decomposition for the real-time demand for Concord.

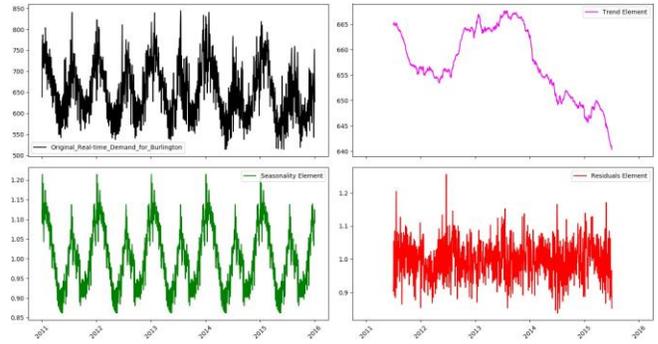

Fig. 9. MD decomposition for the real-time demand for Burlington.

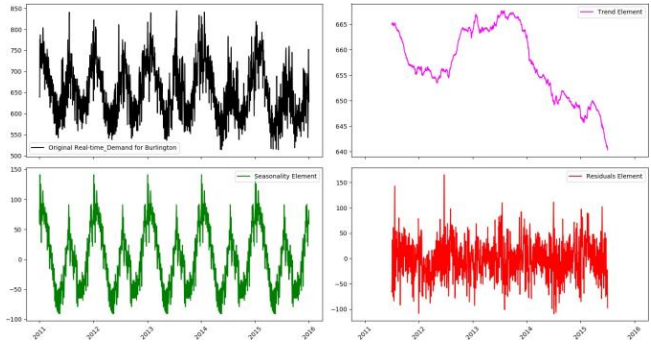

Fig. 8. AD decomposition for the real-time demand for Burlington.

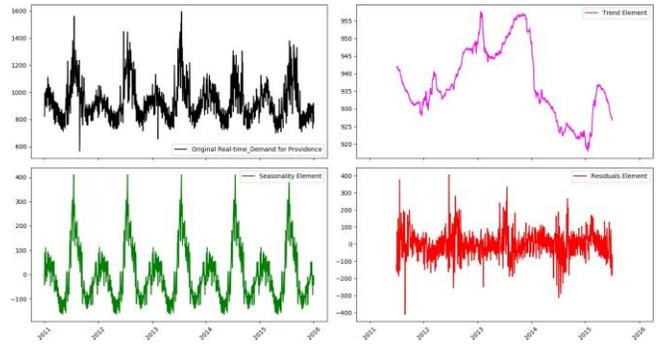

Fig. 10. AD decomposition for the real-time demand for Providence.

### H. Decomposition and Auto-Correlation for Zone 8

The AD and MD decomposition for electricity prices and demands are depicted in Figs. [16]-[17]. The test results for DW and GB test are listed in Table. [8].

TABLE V
PROVIDENCE RT DEMAND PREDICTION ERROR

| RTP Prediction Error | | | | |
|---|---|---|---|---|
| **coeff.** | **std err.** | **t** | **[0.025** | **0.975]** |
| 0.1605 | 0.023 | 6.942 | 0.115 | 0.206 |
| **F-statistic** | 48.19 | **BIC** | 2.073e+04 | |
| **R-squared** | 0.026 | **Adj. R-squared** | 0.025 | |
| **Log-Likelihood** | -10360.3 | **AIC** | 2.072e+04 | |
| **Prob (Omnibus)** | 0.000 | **Skew** | -0.078 | |
| **Jarque-Bera (JB)** | 1043.952 | **Prob (JB)** | 2.04e-227 | |
| **Prob (F-statistic)** | 5.36e-12 | **Omnibus** | 163.949 | |
| **Kurtosis** | 6.702 | **Cond. No.** | 1.00 | |

TABLE VI
SEMASS RT DEMAND PREDICTION ERROR

| RTP Prediction Error | | | | |
|---|---|---|---|---|
| **coeff.** | **std err.** | **t** | **[0.025** | **0.975]** |
| 0.1750 | 0.023 | 7.591 | 0.130 | 0.220 |
| **F-statistic** | 57.63 | **BIC** | 2.272e+04 | |
| **R-squared** | 0.031 | **Adj. R-squared** | 0.030 | |
| **Log-Likelihood** | -11357 | **AIC** | 2.272e+04 | |
| **Prob (Omnibus)** | 0.000 | **Skew** | -0.138 | |
| **Jarque-Bera (JB)** | 750.482 | **Prob (JB)** | 1.08e-163 | |
| **Prob (F-statistic)** | 5.03e-14 | **Omnibus** | 144.916 | |
| **Kurtosis** | 6.129 | **Cond. No.** | 1.00 | |

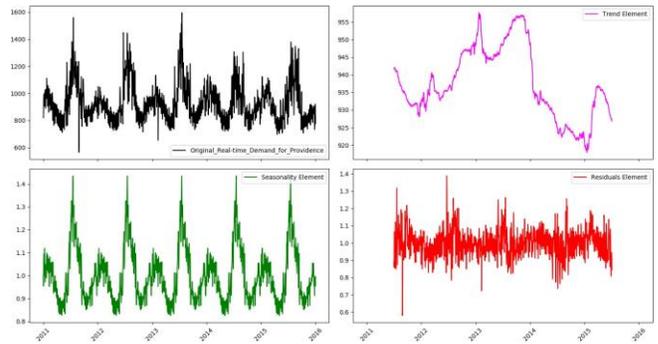

Fig. 11. MD decomposition for the real-time demand for Providence.

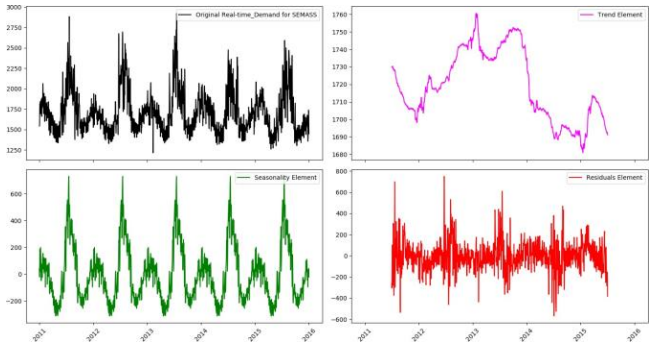

Fig. 12. AD decomposition for the real-time demand for SEMASS.

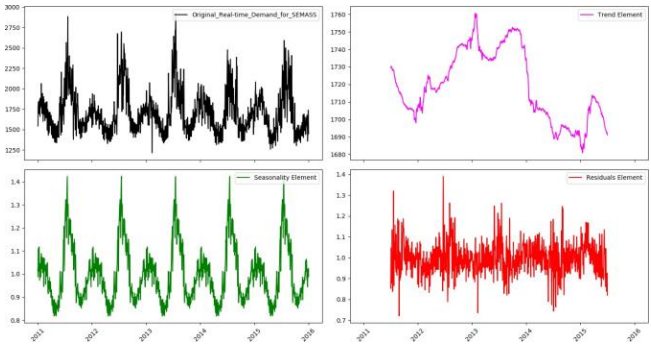

Fig. 13. MD decomposition for the real-time demand for SEMASS.

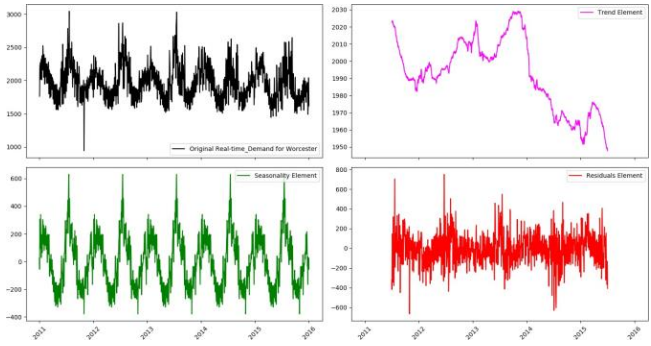

Fig. 14. AD decomposition for the real-time demand for Worcester.

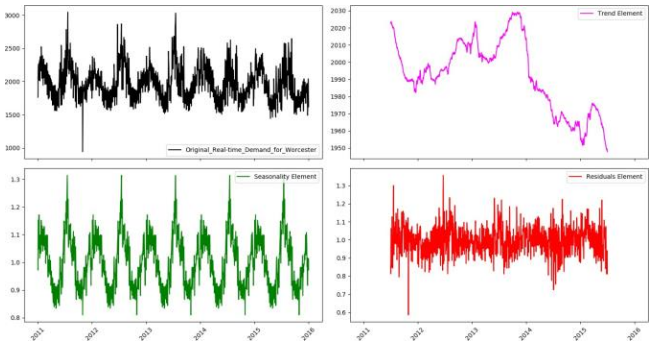

Fig. 15. MD decomposition for the real-time demand for Worcester.

TABLE VII
WORCESTER RT DEMAND PREDICTION ERROR

| RTP Prediction Error | | | | |
|---|---|---|---|---|
| coeff. | std err. | t | [0.025 | 0.975] |
| 0.1817 | 0.023 | 7.887 | 0.137 | 0.227 |
| **F-statistic** | 62.20 | **BIC** | 2.341e+04 | |
| **R-squared** | 0.033 | **Adj. R-squared** | 0.032 | |
| **Log-Likelihood** | -11702 | **AIC** | 2.341e+04 | |
| **Prob (Omnibus)** | 0.000 | **Skew** | 0.255 | |
| **Jarque-Bera (JB)** | 190.651 | **Prob (JB)** | 1.08e-163 | |
| **Prob (F-statistic)** | 5.30e-15 | **Omnibus** | 81.973 | |
| **Kurtosis** | 4.499 | **Cond. No.** | 1.00 | |

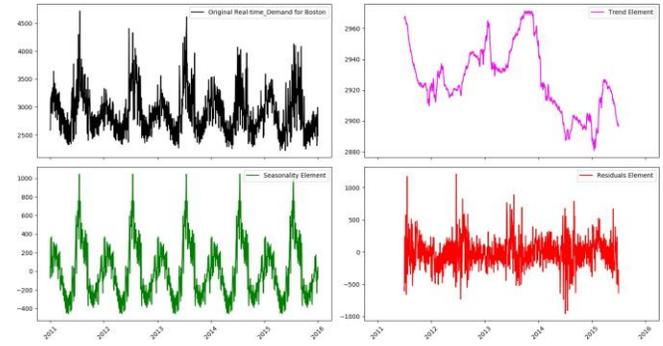

Fig. 16. AD decomposition for the real-time demand for Boston.

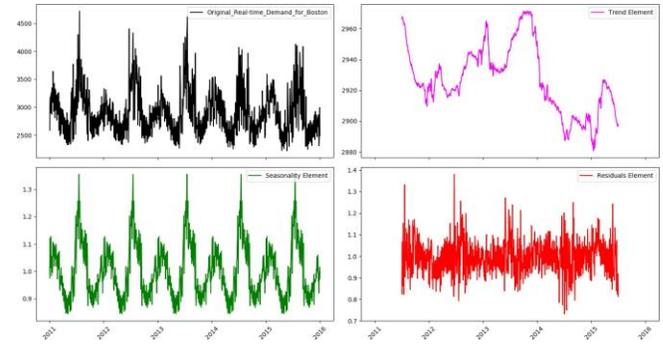

Fig. 17. MD decomposition for the real-time demand for Boston.

TABLE VIII
BOSTON RT DEMAND PREDICTION ERROR

| RTP Prediction Error | | | | |
|---|---|---|---|---|
| coeff. | std err. | t | [0.025 | 0.975] |
| 0.1706 | 0.023 | 7.392 | 0.125 | 0.216 |
| **F-statistic** | 54.65 | **BIC** | 2.487e+04 | |
| **R-squared** | 0.029 | **Adj. R-squared** | 0.029 | |
| **Log-Likelihood** | -12429 | **AIC** | 2.486e+04 | |
| **Prob (Omnibus)** | 0.000 | **Skew** | 0.152 | |
| **Jarque-Bera (JB)** | 264.073 | **Prob (JB)** | 4.54e-58 | |
| **Prob (F-statistic)** | 2.19e-13 | **Omnibus** | 86.875 | |
| **Kurtosis** | 4.839 | **Cond. No.** | 1.00 | |